\documentclass{article}
\usepackage{a4}
\usepackage{epsfig}
\usepackage{amsfonts}
\usepackage{amsmath}
\usepackage{amsthm}
\usepackage{amssymb}

\title{Second order reductions of the WDVV Equations related to classical Lie algebras}
\author{L.K. Hoevenaars \and R. Martini}

\newtheorem{proposition}{Proposition}[section]

\begin{document}
\date{}
\maketitle

\begin{abstract}
\noindent
We construct second order reductions of the generalized Witten-Dijkgraaf-Verlinde-Verlinde system based on simple Lie algebras. We discuss to what extent some of the symmetries of the WDVV system are preserved by the reduction.
\end{abstract}

{\noindent}\textbf{MSC Subj. Class. 2000: }35C05, 81T60

{\noindent}\textbf{Keywords: }WDVV equations, prepotentials, Seiberg-Witten theory

\section{Introduction} 
In two-dimensional topological conformal field theory the following remarkable system of third order nonlinear partial differential equations for a function $F$ of $N$ variables emerged \cite{WITT:1991},\cite{DIJK-VERL-VERL:1991}
\begin{equation}\label{eq1}
F_i F_1^{-1} F_j = F_j F^{-1}_1 F_i \qquad i,j = 1, \ldots , N
\end{equation}
Here $F_i$ is the matrix
\[
(F_i)_{kl} = \frac{\partial^3 F}{\partial x_i \partial x_k \partial x_l}\;.
\]
Roughly speaking, this system expresses the conditions on a function whose third order derivatives form the structure constants of an associative, commutative algebra.
Moreover, it is required that $F_1$ is a constant and invertible matrix. Usually this system is called the Witten-Dijkgraaf-Verlinde-Verlinde (WDVV) equations. Generalizations, not requiring $F_1$ to be constant, have been introduced and studied in the context of four- and five-dimensional $N=2$ supersymmetric gauge theory (see e.g. \cite{MARS-MIRO-MORO:1996,MARS-MIRO-MORO:2000}, \cite{HOEV-MART2:2003}.

Although extremely difficult to solve in general, this overdetermined system (\ref{eq1}) of nonlinear partial differential equations admits exact solutions. For instance, within the theory of Frobenius manifolds, a substantial class of polynomial solutions has been constructed by Dubrovin \cite{DUBR:1996} for any Coxeter group. Furthermore, for any gauge group, perturbative approximations to exact prepotentials in four-dimensional Seiberg-Witten theory satisfy the (generalized) WDVV equations. These solutions are of logarithmic type and may be constructed for any root system (see \cite{MARS-MIRO-MORO:1996} and \cite{MART-GRAG:1999}). The corresponding five-dimensional theories give rise to trigonometric solutions, see e.g. \cite{NEKR:1998},\cite{HOEV-MART2:2003}.

Inspired by the paper \cite{BRAD-MARS:2002} we construct second order reductions of the (third order) WDVV system, related to the logarithmic solutions associated with classical Lie algebras. In \cite{BRAD-MARS:2002}, the authors' main result is a form of the WDVV equations resembling a cocycle condition
\begin{eqnarray}
\label{braden marshakov}
\sum_{i,k} \frac{\partial T_{ai}}{\partial T_{bj}}\frac{\partial T_{bk}}{\partial T_{ci}}\frac{\partial T_{cl}}{\partial T_{ak}}=\delta^{j}_{l}
\end{eqnarray}
where $T$ is the Hesse matrix of a function $F(x_1,...,x_N)$.
The main idea is to perform a change of variables from $x_k$ to a single row $T_a=\left\{ T_{ak} \right\}$ of the Hesse matrix. All other elements of the Hesse matrix are functionally dependent on that row and the authors then show that the equations (\ref{braden marshakov}) are equivalent to the generalized WDVV system. 

In fact, Braden and Marshakov go on to construct a second order reduction\footnote{A reduction of a system of differential equations for a single function $F$ is understood here to be a set of lower order equations for $F$ whose solutions are automatically solutions of the original system.} of the WDVV equations by studying the relations among the elements of the Hessian of the prepotential of type $A_3$. In this paper, we modify and largely extend the construction of such reductions to all Lie algebras of classical type in section \ref{section reductions}. Finally, we discuss the effect of changes of the ultraviolet cut-off on the reductions in section \ref{section Lambda}.

%It should be noted that whereas (\ref{braden marshakov}) is equivalent to the WDVV equations and therefore shares the same symmetries, the second order reductions do not. We study to what extent some known symmetries of the WDVV system are preserved by the reduction in section \ref{section symmetries}.

%One of the main goals of constructing reductions of partial differential equations is to explicitly solve the reduction, thus gaining new solutions to the original equations. In the course of this paper, we will explain that for the reductions studied here unfortunately no new solutions to the WDVV equations can be obtained in this way.

\section{Second order reductions for classical Lie algebras} %2
\label{section reductions}
For any rank $N$ root system $R$ with positive roots $R^+$, we consider the following function
\begin{eqnarray}
\label{prepotential root systems}
{\cal F}(x_1,...,x_N) = \frac{1}{2} \sum_{\alpha \in R^+} (\alpha,x)^2 \ln \left(\frac{(\alpha,x)}{\Lambda}\right)
\end{eqnarray}
where $x=\sum_i x_i v_i$ is a linear combination of basis vectors $v_i$ of the root space, and $(.,.)$ denotes the Euclidean inner product. 
Such functions are known as pure perturbative four-dimensional Seiberg-Witten prepotentials in the physics literature \cite{SEIB-WITT1:1994,KLEM-LERC-YANK-THEI:1995}, where they occur in the study of quantum Yang-Mills theories.
The parameter $\Lambda$ occurs in the Seiberg-Witten theory as an energy cut-off scale, and is irrelevant for the WDVV equations since it has no effect on the third order derivatives of ${\cal F}$. The reductions we propose here are of second order, but as we will show in section \ref{section Lambda} they turn out also to be independent of the value of $\Lambda$.

\subsection{The reduction for type $A$ prepotentials}
Naturally, the explicit form of the prepotential depends on the chosen basis $\{v_i\}$. Although this choice is usually irrelevant since linear transformations are symmetries of the WDVV equations, we will be dealing with second order reductions on which the choice may have an influence. The choice we make here leads to the well-known type $A$ Seiberg-Witten prepotential
\begin{eqnarray}
\label{prepotential A}
{\cal F}(x_1,...,x_N)=\frac{1}{2} \sum_{i<j} (x_i-x_j)^2\ln \left(\frac{x_i-x_j}{\Lambda} \right) + \frac{1}{2}\sum_i x_i^2\ln\left( \frac{x_i}{\Lambda} \right)
\end{eqnarray}
For convenience we set $\Lambda=e^{\frac{3}{2}}$, after which the second order derivatives $T_{ij}$ of ${\cal F}$ are given by
\begin{eqnarray}
e^{-T_{ij}} &=& x_i-x_j \qquad i<j
\nonumber \\
e^{T_{ii}} &=& x_i \prod_{q>i} (x_i-x_q) \prod_{q<i} (x_q - x_i)
\end{eqnarray}
These second order derivatives satisfy the following set of relations
\begin{eqnarray}
\label{relaties1}
e^{-T_{ik}} =& e^{-T_{ij}} + e^{-T_{jk}} & \forall i<j<k \\
e^{-T_{ij}} =& e^{\sum_p T_{ip}} - e^{\sum_q T_{jq}} & \forall i<j
\label{relaties2}
\end{eqnarray}
These relations are somewhat reminiscent of those occurring in Hirota's work on tau functions of integrable hierarchies \cite{BRAD-MARS:2002}, \cite{ZABR:1997}. 

%We denote by ${\cal F}_i$ the matrix with indices $\left({\cal F}_i \right)_{jk}$ the third order derivatives of ${\cal F}$ with respect to $x_i,x_j,x_k$. 
%The WDVV equations require the existence of an invertible linear combination of the ${\cal F}_a$, and in the present case this condition is fulfilled since any $F_a$ is itself invertible for the prepotential (\ref{prepotential A}). This implies that the change of coordinates from $x_k$ to $T_{ak}$ is a good one, i.e. its Jacobi matrix is nondegenerate. From the relations (\ref{relaties1}),(\ref{relaties2}) we can find the explicit change of coordinates
Starting from the relations (\ref{relaties1}),(\ref{relaties2}) one sees that the elements $T_{ij}$ of the Hesse matrix $T$ can be expressed in terms of a single row $T_a=\left\{ T_{ak} \right\}$ for fixed $a$ as follows
\begin{eqnarray}
T_{ij} = -\log \left(e^{-T_{ia}}+e^{-T_{aj}} \right) \qquad i<a<j
\nonumber \\
T_{ij} = -\log \left(e^{-T_{aj}}- e^{-T_{ai}} \right) \qquad a<i<j
\nonumber \\
T_{ij} = -\log \left(e^{-T_{ia}}- e^{-T_{ja}} \right) \qquad i<j<a
\end{eqnarray}
and
\begin{eqnarray}
T_{ii} =& \log \left(  e^{-\sum_{p \neq i}T_{ip}}  \left[e^{T_{ij}} + e^{\sum_q T_{aq}}\right]        \right) &  \qquad i<a
\nonumber \\
T_{ii} =& \log \left(  e^{-\sum_{p \neq i}T_{ip}}  \left[-e^{T_{ij}} + e^{\sum_q T_{aq}}\right]  \right) & \qquad i>a
\end{eqnarray}
In general no relations exist between the $T_{ak}$ and they can be taken as independent.

%We thus have the following lemma
%\begin{lemma}
%For any function whose second order derivatives satisfy the relations (\ref{relaties1}),(\ref{relaties2}) one can make a change of coordinates from $x_k$ to $T_{ik}$ for any fixed $i$.
%\end{lemma}

\begin{proposition}
\label{prop A type}
The relations (\ref{relaties1}),(\ref{relaties2}) form a second order reduction of the WDVV equations in the sense that any function $G(z_1,...,z_N)$ whose second order derivatives $T_{ij}$ satisfy the reduction, automatically satisfies the WDVV equations. 
\end{proposition}
\begin{proof}
Having a function $G(z)$ satisfying the reduction, we use the equations
\begin{eqnarray}
e^{-T_{ij}} &=& x_i-x_j  \qquad \qquad \qquad \qquad \qquad i<j
\nonumber \\
e^{T_{ii}} &=& x_i \prod_{q>i} (x_i-x_q) \prod_{q<i} (x_q - x_i)
\label{introductie x}
\end{eqnarray}
to introduce the objects $x_i(z)$. We already know from the study of the prepotential (\ref{prepotential A}) that these equations are compatible with the reduction. Moreover it is easily seen that (\ref{introductie x}) can be used to define the $x_i$ uniquely. Since the objects $\frac{\partial T_{ij}}{x_k}$ are totally symmetric, the $T_{ij}$ can be integrated twice with respect to the $x_i$ (not the $z_i$) to obtain the known prepotential (\ref{prepotential A}). We now have two functions, the well-known ${\cal F}(x)$ and the unknown $G(z)$, whose second order derivatives are equal $\frac{\partial^2 {\cal F}}{\partial x_i \partial x_j}(x(z))=\frac{\partial^2 G}{\partial z_i \partial z_j}(z)$.
As shown in \cite{MIRO-MORO:1998}, due to this relation between the second order derivatives the unknown function $G(z)$ satisfies the WDVV equations since the prepotential ${\cal F}$ does.
\end{proof}

\subsection{The reduction for the remaining classical Lie algebras}
The roots in $\mathbb{R}^N$ of the remaining classical Lie algebras are given in table \ref{table BCD}.
Taking the standard Euclidean basis, the corresponding prepotentials are
\begin{eqnarray}
\label{prepotential BCD}
{\cal F}(x_1,...,x_N) = \frac{1}{2} \sum_{i<j} 
\left( 
(x_i-x_j)^2\ln \left(\frac{x_i-x_j}{\Lambda} \right) 
+
(x_i+x_j)^2\ln \left(\frac{x_i+x_j}{\Lambda} \right)
\right) 
\nonumber \\
+ \eta \sum_i x_i^2\ln\left( \frac{x_i}{\Lambda} \right)
+ \zeta  \sum_i (x_i)^2
\end{eqnarray}
with $\eta,\zeta$ as in table \ref{table BCD}. 
Chosing $\Lambda=e^{\frac{6(N-1+\eta)+4\zeta}{4(N-1+\eta)}}$, one finds
\begin{eqnarray}
e^{-T_{ij}}&=&\frac{x_i-x_j}{x_i+x_j}
\\
e^{T_{ii}} &=& x_i^{2\eta} \prod_{q>i} (x_i^2-x_q^2) \prod_{q<i} (x_q^2 - x_i^2)
\end{eqnarray}
From these values of the second order derivatives of the prepotential one can derive the following relations among them
\begin{eqnarray}
\label{relaties3}
\coth \left(\frac{T_{ik}}{2} \right) =& \hspace{-1.7cm} \coth \left(\frac{T_{ij}}{2} \right)\coth \left(\frac{T_{jk}}{2} \right) & \quad \forall i<j<k
\\
e^{T_{ii}-T_{jj}} =& \left[ \coth \left(\frac{T_{ij}}{2}\right) \right]^{2\eta+N-2}\frac{\prod_{n\neq j}\sinh(T_{jn})}{\prod_{m\neq i}\sinh(T_{im})}
& \quad \forall i<j
\label{relaties4}
\end{eqnarray}
We therefore have the following result
\begin{proposition}
\label{prop BCD type}
The relations (\ref{relaties3}),(\ref{relaties4}) form a second order reduction of the WDVV equations in the sense that any function $G(z_1,...,z_N)$ whose second order derivatives $T_{ij}$ satisfy the reduction, also satisfies the WDVV equations. 
\end{proposition}
\begin{proof}
The proof follows the same lines as the proof of proposition \ref{prop A type}.
\end{proof}

\begin{center}
\begin{table}
\centering
\begin{tabular}{|l|l|l|}
\hline
$B_N$ & $\pm e_i \pm e_j, e_i$  & $\eta =\frac{1}{2}, \zeta =0$
\\
\hline
$C_N$ & $\pm e_i \pm e_j, 2e_i$ & $\eta =2, \zeta =2\ln(2)$
\\
\hline
$D_N$ & $\pm e_i \pm e_j$ & $\eta=\zeta=0$
\\
\hline
\end{tabular}
\caption{\emph{The roots of the type $B,C,D$ Lie algebras together with the values of $\eta,\zeta$ occurring in the prepotential (\ref{prepotential BCD}). Here $e_i$ denotes a vector with $1$ in the i-th place and $0$ elsewhere. }}
\label{table BCD}
\end{table}
\end{center}

\section{The role of $\Lambda$}
\label{section Lambda}
The parameter $\Lambda$ plays the role of the energy cut-off in Seiberg-Witten theory. Changing this parameter amounts to adding a second order polynomial to the prepotential, which is clearly a symmetry of the third order WDVV equations. 
The reductions proposed in this paper are of second order, and so the value of $\Lambda$ can in principal make a difference. We will show however that this is not the case.
\begin{proposition}
Adding a second order polynomial preserves the solution set of the reduction $(\ref{relaties1}),(\ref{relaties2})$ iff the polynomial is a multiple of $\left( \sum_{i<j}(x_i-x_j)^2 + \sum_i x_i^2 \right)$. Similarly, addition of a second order polynomial preserves the solution set of the reduction $(\ref{relaties3}),(\ref{relaties4})$ iff it is of the form $\sum_{i \neq j}n\pi \sqrt{-1}  x_ix_j+\xi \sum_i x_i^2$ with arbitrary integer $n$ and arbitrary parameter $\xi$.
\end{proposition}
\begin{proof}
Adding a polynomial $\sum_{i,j}a_{ij}x_ix_j$ to a generic solution of the type $A_N$ reduction, one finds that (\ref{relaties1}) restricts $a_{ij}=c$ for $i\neq j$ and some common constant $c$. Then it is clear that (\ref{relaties2}) restricts $a_{ii}=-Nc$. This leads to the assertion of the lemma regarding the type $A$ reduction. Similarly (\ref{relaties3}) restricts the polynomial to $\sum_{i < j}n\pi \sqrt{-1} x_ix_j+\sum_i a_{ii}x_i^2$, and (\ref{relaties4}) restricts the $a_{ii}$ further.
\end{proof}
A change $\Lambda \rightarrow e^{-2\xi}\Lambda$ in (\ref{prepotential A}) leads to an addition of a term $\xi\left( \sum_{i<j}(x_i-x_j)^2 + \sum_i x_i^2 \right)$ which is a symmetry of the type $A$ reduction. Similarly, a change $\Lambda \rightarrow e^{-\frac{\xi}{N-1+\eta}}\Lambda$ in (\ref{prepotential BCD}) gives an additional term $\xi \sum_i x_i^2$ which in turn is a symmetry of the $BCD$ reduction.

\section{Remarks}
The generalized WDVV equations are known to be invariant under the following two types of transformations \cite{DUBR:1996,MARS-WIT:2001}: linear transformations of the coordinates and so-called duality transformations. The duality transformations comprise a group of contact symmetries acting as constant symplectic transformations on a vector containing the coordinates $x_i$ as one half of its indices and the first order derivatives $F_i$ of the prepotential as the other half.
It is readily observed that the reductions (\ref{relaties1}), (\ref{relaties2}) and (\ref{relaties3}), (\ref{relaties4}) are not invariant under general linear coordinate changes nor general symplectic transformations. This implies among other things that the choice of basis in the root space is relevant for the reduction one obtains.

A purpose of having a reduction might be to see if it can be easily solved, thus gaining new solutions to the WDVV equations. 
As shown above, the solutions of the reductions consist of the functions $G(z)$ whose Hesse matrix equals that of the perturbative prepotentials ${\cal F}(x)$ for classical Lie algebras: $\frac{\partial^2 G}{\partial z_i \partial z_j}(z)=\frac{\partial^2 F}{\partial x_i \partial x_j}(x(z))$.
Since this type of relation between two prepotentials was already studied in the literature (see e.g. \cite{MIRO-MORO:1998}) the reduction does not yield new information in this sense.

\bibliographystyle{abbrv}
\bibliography{../../biblio/biblio}
\include{thebibliography}

\end{document}